\documentclass[aps,prd,preprint,superscriptaddress,amsmath,amssymb,showpacs]{revtex4-1}
\usepackage{dcolumn}
\usepackage{graphicx}
\usepackage{float}
\usepackage{physics}
\usepackage[colorlinks=true,allcolors=blue]{hyperref}
\begin{document}
	\title{Chiral Phase Transition in Rotating Quark Matter with Chiral Imbalance:
A Medium Separation Scheme Regularized NJL Model Study
}
	
	\author{Huang-Jing Zheng}
    \altaffiliation{These authors contributed equally to this work.}
	\affiliation{College of Mathematics and Physics, China Three Gorges University, Yichang 443002, China}

	\author{Peng Nan}
    \altaffiliation{These authors contributed equally to this work.}
	\affiliation{College of Mathematics and Physics, China Three Gorges University, Yichang 443002, China}

	\author{Sheng-Qin Feng}
	\email{Corresponding author: fengsq@ctgu.edu.cn}
	\affiliation{College of Mathematics and Physics, China Three Gorges University, Yichang 443002, China}
	\affiliation{Center for Astronomy and Space Sciences and Institute of Modern Physics, Three Gorges
University, Yichang 443002, China
}
	\affiliation{Key Laboratory of Quark and Lepton Physics (MOE) and Institute of Particle Physics, Central China Normal University, Wuhan 430079, China}
	
	\date{\today}

	\begin{abstract}
		We investigate the chiral phase transition in rotating quark matter with chiral imbalance using the two-flavor Nambu-Jona-Lasinio (NJL) model regularized by the Medium Separation Scheme (MSS). Our numerical calculations demonstrate that the chiral chemical potential $\mu_5$ and angular velocity $\omega$ exert opposite effects on chiral symmetry breaking: $\mu_5$ enhances the breaking, raising the pseudocritical temperature $T_{pc}$ and sharpening the phase transition, while $\omega$ suppresses the breaking, lowering $T_{pc}$ and smearing the transition. Notably, chiral imbalance buffers the rotation-induced softening of the phase transition-the suppression of $T_{pc}$ by $\omega$ weakens progressively as $\mu_5$ increases. The MSS predicts a monotonic increase of $T_{pc}$ with $\mu_5$, in qualitative agreement with LQCD, resolving the discrepancy found in traditional regularization. Furthermore, the rotational suppression of $T_{pc}$ exhibits strong radius dependence: larger rotation radii amplify the suppression due to enhanced spacetime curvature and centrifugal effects, and can even induce an abrupt drop in $T_{pc}$ in the high-rotation region. These findings clarify the interplay between rotation and chiral imbalance in modulating the QCD chiral phase transition and validate the MSS as a reliable regularization framework for such extreme systems.
	\end{abstract}
	
	\maketitle
	
\section{Introduction}\label{sec:Intro}
In recent years, the role of rotation in quantum chromodynamics (QCD) phase transitions has emerged as a prominent research topic. Non-central heavy-ion collisions not only generate hot and dense quark-gluon plasma (QGP) but also impart enormous orbital angular momentum, reaching magnitudes of the order of 10$^4\hbar$~\cite{1,2}, corresponding to local angular velocities $\omega\sim$  0.1 GeV \cite{1,2,3,4,5,6,7,8}. Such extreme rotational conditions profoundly influence the spin and orbital angular momentum dynamics of quarks and gluons \cite{9,10,11,12,13,14}, offering a unique opportunity to explore the properties of the QCD vacuum structure and the phase diagram of strongly interacting matter under the influence of rotation. Spin-orbit coupling in QCD can induce a variety of interesting phenomena. For instance, the Beam Energy Scan (BES) program at the STAR Collaboration has systematically measured global polarization in Au-Au non-central collisions, providing the first experimental evidence for such an extreme rotational environment \cite{11}. This discovery opens new avenues for investigating the QCD phase diagram under rotation, particularly regarding vorticity - induced topological effects and chiral imbalance. Recent theoretical studies have further explored the impact of rotation on chiral symmetry breaking and confinement - deconfinement transition using effective models and lattice simulations \cite{12,13,14,15,16}.

Parallel to the interest in rotational effects, the study of how chiral imbalance between right- and left-handed quarks affects the QCD phase diagram has attracted increasing attention. The non-trivial topological structure of the QCD vacuum, characterized by instantons \cite{17} and sphalerons \cite{18,19}, gives rise to configurations with non-zero topological winding number $Q_W$. Through the axial anomaly, these topological fluctuations induce local changes in the vacuum topological charge and flip quark helicities, thereby generating a chiral imbalance i.e., a difference between the number densities of left- and right-handed quarks \cite{20,21,22,23,24}. This process leads to the violation of parity ($P$) and charge-parity ($CP$) symmetry in the thermal plasma \cite{22,23,24}. The interplay between chiral imbalance and other external conditions, such as magnetic fields and rotation, has become a subject of intense investigation \cite{25,26,27,28}.

To quantify this chiral imbalance, a chiral chemical potential $\mu_5$ conjugate to the chiral charge is introduced. Within the grand canonical ensemble, the effects of chiral imbalance on the QCD phase diagram can be systematically studied by incorporating $\mu_5$ \cite{20}. Beyond the intrinsic motivation related to heavy-ion collision physics, recent studies suggest that the QCD phase diagram in the $T -\mu_5$ plane could, in principle, be mapped onto the real phase diagram in the $T -\mu$ plane (where $\mu $ is the quark chemical potential) \cite{29, 30}. More importantly, QCD in the presence of a chiral chemical potential is free from the sign problem, making lattice simulations with $\mu_5$ a valuable benchmark platform for comparing different effective models. However, a significant discrepancy has emerged: while traditionally regularized Nambu-Jona-Lasinio (NJL) - type models \cite{21,22,24,31,32} and quark linear sigma models \cite{20,33} predict that the critical temperature $T_c$ decreases with increasing $\mu_5$, lattice QCD results \cite{34, 35} show that $T_c$ increases with $\mu_5$. This contradiction highlights the need for a more careful regularization approach.

The NJL model is non-renormalizable, and therefore its regularization must be handled with caution. In principle, the regularization procedure should act only on vacuum contributions, while medium-dependent terms should remain unaffected by the chosen regularization scheme. This rigorous separation between medium effects and the regularized vacuum component is not only qualitatively important but also has significant quantitative implications. This conclusion has been confirmed in studies of color superconductivity \cite{36}. A similar situation arises in the study of magnetized quark matter: by appropriately separating the magnetic field contribution from divergent integrals \cite{37,38,39,40,41}, unphysical artifacts can be effectively eliminated. Recent investigations of chiral imbalanced NJL models \cite{42} also advocate for the same treatment. This approach, which decouples the vacuum term from the medium-dependent part and applies regularization only to the ultraviolet divergent momentum integrals of the former, has been termed the ``medium separation scheme" (MSS), in contrast to the traditional regularization scheme (TRS) where the cutoff is applied directly to medium-dependent terms as well.

Since its development in 2006, MSS has been widely applied in the study of quark matter under various extreme conditions, including color superconductivity \cite{36, 43}, neutron star matter \cite{44}, and quark matter with chiral imbalance \cite{42,45,46}. Previous studies have verified that MSS effectively eliminates the spurious cutoff dependence of medium contributions inherent in TRS, and its predictions for key physical quantities such as the pseudocritical temperature of the chiral phase transition are in good qualitative agreement with LQCD results \cite{32,45,47,48}. The success of MSS in describing magnetized and chiral-imbalanced systems suggests its potential applicability to rotating quark matter, where similar regularization issues may arise \cite{49, 50}. In this work, considering a rotating quark matter system with chiral imbalance, we adopt the MSS to regularize the NJL model, aiming to accurately describe the modulatory effects of rotation and chiral chemical potential on the chiral phase transition and to reveal the intrinsic mechanism of their interplay.

The structure of this paper is as follows. In Sec. II, we introduce the two-flavor NJL model with chiral imbalance under rotation and present the MSS regularization procedure. In Sec. III, we numerically investigate the detailed influences of chiral imbalance and rotation on the chiral phase transition. Finally, a summary and conclusions are given in Sec. IV.
\section{NJL MODEL REGULARIZED BY THE MSS WITH CHIRAL IMBALANCE
UNDER ROTATION
}\label{sec:2}
Let's consider a system composed of spin particles. In a reference frame of co-rotating with the system at a constant angular velocity~$\omega $, the originally flat Minkowski spacetime acquires an effective curvature metric. The local static observers in this reference frame measure a local velocity $\vec{v}=\vec{\omega }\,\ \times \ \vec{r}$. The line element of the metric induced by the rotation coordinates \cite{15} takes the form as
\begin{equation}\label{eq:1}
\left. {{g}_{\mu \nu }}=\left( \begin{matrix}
   1-{{{\vec{v}}}^{2}} & -{{v}_{1}} & -{{v}_{2}} & -{{v}_{3}}  \\
   -{{v}_{1}} & -1 & 0 & 0  \\
   -{{v}_{2}} & 0 & -1 & 0  \\
   -{{v}_{3}} & 0 & 0 & -1  \\
\end{matrix} \right. \right).
\end{equation}

The Lagrangian of the two-flavour NJL model in this setting is given as
\begin{equation}
    {{\mathcal{L}}_{QCD}}=\bar{\psi }[i{{\bar{\gamma }}^{\mu }}{{D}_{\mu }}-m]\psi +{{G}_{s}}[{{(\bar{\psi }\psi )}^{2}}+{{(\bar{\psi }i{{\gamma }^{5}}\vec{\tau }\psi )}^{2}}],
\end{equation}
where ${{D}_{\mu }}={{\partial }_{\mu }}+{{\Gamma}_{\mu }}$ is the covariant derivative in the rotating frame, $\psi $ denotes the quark fields for the two light flavours $u$ and $d$, $m$ is the fermion mass, ${{G}_{s}}$ is the coupling constant in the scalar-pseudoscalar channel. As one defines${{\bar{\gamma }}^{\mu }}=e_{a}^{\mu }{{\gamma }^{a}}$, where$e_{a}^{\mu }$ is the tetrads field for spinor fields and ${{\gamma }^{a}}$ are the usual Dirac matrices in the local Lorentz frame. The spin connection is given by ${{\Gamma}_{\mu }}=\frac{1}{4}\times \frac{1}{2}[{{\gamma }^{a}},{{\gamma }^{b}}]{{\Gamma}_{ab\mu }}$, with ${{\Gamma}_{ab\mu }}={{\eta }_{ac}}\left( e_{\sigma }^{c}G_{\mu \nu }^{\sigma }e_{b}^{\nu }-e_{\nu }^{b}{{\partial }_{\mu }}e_{\nu }^{c} \right)$, where $G_{\mu \nu }^{\sigma }$ is the affine connection determined by the metric${{g}_{\mu \nu }}$. The simplest tetrad choice: $e_{\mu }^{a}=\delta _{\mu }^{a}+\delta _{i}^{a}\delta _{\mu }^{0}{{v}^{i}}$, $e_{a}^{\mu }=\delta _{a}^{\mu }-\delta _{a}^{0}\delta _{i}^{\mu }{{v}^{i}}$ is employed. We introduce a chiral chemical potential $\mu_5$ to simulate the chiral imbalance between left- and right-handed quarks, and the Lagrangian density can be given as $\mathcal{L}={{\mathcal{L}}_{\mathcal{Q}\mathcal{C}\mathcal{D}}}+{{\mu }_{5}}{{\psi }^{\dagger }}{{\gamma }^{5}}\psi $ \cite{20}. The Lagrangian density of the two-flavor NJL model with chiral imbalance under rotation can be obtained as
\begin{equation}
 {{\mathcal{L}}_{\text{NJL}}}=\bar{\psi }[i{{\bar{\gamma }}^{\mu }}({{\partial }_{\mu }}+{{\Gamma}_{\mu }})-m+{{\mu }_{5}}{{\gamma }^{0}}{{\gamma }^{5}}]\psi +{{G}_{s}}[{{(\bar{\psi }\psi )}^{2}}+{{(\bar{\psi }i{{\gamma }^{5}}\vec{\tau }\psi )}^{2}}].
\end{equation}

By using the mean-field approximation, one can obtain the Lagrangian density under rotation \cite{51,52,53} as
\begin{equation}
	{{\mathcal{L}}_{\text{MFA}}}=\bar{\psi }[i{{\gamma }^{\mu }}{{\partial }_{\mu }}-M+{{\mu }_{5}}{{\gamma }^{0}}{{\gamma }^{5}}+{{({{\gamma }^{0}})}^{-1}}(\vec{\omega }\times \vec{r})\cdot (-i\vec{\partial })+\vec{\omega }\cdot \vec{S})]\psi +\frac{{{\sigma }^{2}}}{4{{G}_{s}}},	
    \end{equation}
where $\sigma =-2{{G}_{s}}\langle \bar{\psi }\psi \rangle $ is the chiral condensate, $M = m +\sigma $ is the dynamical quark mass \cite{54,55,56,57,58}, $\omega $ is the angular velocity of rotation, and the spin operator is given as
\begin{equation}
	{{\vec{S}}_{4\times 4}}=\frac{1}{2}\left( \begin{matrix}
   {\vec{\sigma }} & 0  \\
   0 & {\vec{\sigma }}  \\
\end{matrix} \right).	
\end{equation}

By using cylindrical coordinates, one can derive the general positive-energy solution of the quark field from the Dirac equation corresponding to the above Lagrangian as
	\begin{equation}\psi (\theta ,r)={{e}^{-iEt+i{{p}_{z}}z}}\left( \begin{matrix}
   c{{e}^{in\theta }}{{J}_{n}}({{p}_{t}}r)  \\
   id{{e}^{i(n+1)\theta }}{{J}_{n+1}}({{p}_{t}}r)  \\
   {c}'{{e}^{in\theta }}{{J}_{n}}({{p}_{t}}r)  \\
   i{d}'{{e}^{i(n+1)\theta }}{{J}_{n+1}}({{p}_{t}}r)  \\
\end{matrix} \right),	
\end{equation}
where ${{J}_{n}}({{p}_{t}}r)$ is a Bessel function of the first kind. The coefficients $c, d,{c}',$ and ${d}'$ satisfy the normalization condition ${{c}^{2}}+{{d}^{2}}+{{c}^{\prime 2}}+{{d}^{\prime 2}}=1$, and they also satisfy $d/c=-{d}'/{c}'=\lambda $, where $\lambda $ is expressed as
	\begin{equation}\lambda =\frac{-2{{\mu }_{5}}{{p}_{t}}}{{{\left( \sqrt{p_{t}^{2}+p_{z}^{2}}-s{{\mu }_{5}} \right)}^{2}}-p_{t}^{2}-{{({{p}_{z}}+{{\mu }_{5}})}^{2}}}.
    \end{equation}
The energy of the "free particle" is
${{E}_{p}}=\sqrt{{{(|p|-s{{u}_{5}})}^{2}}+{{M}^{2}}}=\sqrt{{{\left( \sqrt{{{p}_{t}}^{2}+{{p}_{z}}^{2}}-s{{\mu }_{5}} \right)}^{2}}+{{M}^{2}}}$
($s=\pm 1$), in the relativistic quantum system with the "rotation effect", the energy ${{E}_{n,s}}={{E}_{p}}-(n+\frac{1}{2})\omega $. The effective potential is divided into vacuum and thermodynamic parts, and then the ultraviolet components of the vacuum part were regularized by some MSS models \cite{59,60}. The grand potential density can be written as
	\begin{equation}
    \Omega=\frac{{{\sigma }^{2}}}{4{{G}_{s}}}+{{\Omega}_{\text{vac}}}+{{\Omega}_{\text{th}}},
    \end{equation}
    \begin{equation}
                         {{\Omega}_{\text{vac}}}=-2{{N}_{f}}{{N}_{\text{c}}}\int{\frac{{{d}^{3}}p}{{{(2\pi )}^{3}}}}{{E}_{p}},                          \end{equation}
	\begin{equation}{{\Omega}_{\text{th}}}=-\frac{{{N}_{f}}{{N}_{c}}T}{4{{\pi }^{3}}}\sum\limits_{n=-\infty }^{\infty }{\sum\limits_{s=\pm 1}{\int{d{{p}_{t}}^{2}d{{p}_{z}}}}}\frac{[J_{n}^{2}+{{\lambda }^{2}}J_{n+1}^{2}]}{(1+{{\lambda }^{2}})}\ln \left[ 1+{{e}^{\left( -\frac{{{E}_{n,s}}}{T} \right)}} \right],	
    \end{equation}
where ${{N}_{f}}$ and ${{N}_{c}}$ are the numbers of the flavours and colors, which are 2 and 3, respectively. The constituent mass/chiral condensate can be solved by the corresponding gap equations $\frac{\partial \Omega}{\partial \sigma } = 0$ and $\frac{{{\partial }^{2}}\Omega}{\partial {{\sigma }^{2}}}>0$.

Note that the ${{\Omega}_{\text{vac}}}$ in Eq. (9) has ultraviolet divergence, and therefore regularization is required. The traditional regularization scheme (TRS) is a commonly used method in effective field theories such as NJL/PNJL to deal with ultraviolet divergence. By introducing a three-dimensional momentum cutoff $\Lambda $, the divergent integral is restricted to a finite momentum interval, thereby enabling numerical calculation. However, when TRS deals with the explicit contribution of the chiral chemical potential ${{\mu }_{5}}$ to the vacuum momentum integral, some problems may be introduced, leading to results that are inconsistent with the qualitative features predicted by LQCD. Early studies employing the TRS within a variety of effective QCD models \cite{21, 22, 29, 30, 33} reported that the pseudocritical temperature ${{T}_{\textrm{pc}}}$ for chiral symmetry restoration decreases with increasing quark chemical potential ${{\mu }_{5}}$. In contrast, LQCD calculations \cite{34, 35} have reached the opposite conclusion: ${{T}_{\textrm{pc}}}$ increases with ${{\mu }_{5}}$, and this transition is a smooth crossover rather than a first-order phase transition.

To resolve the discrepancy between the TRS and LQCD, the MSS is introduced and developed into a more robust regularization method \cite{35,41,46}. By performing integral operations on divergent propagators, one can separate divergent components from medium-dependent contributions. It can effectively accommodate the effects of chiral chemical potential, partially overcoming the limitations of TRS, and thus reproduce the characteristic behavior of the pseudocritical temperature for chiral symmetry restoration as a function of chiral chemical potential, as predicted by LQCD.

The derivative of the momentum integral with respect to ${{M}^{2}}$ is given as
	\begin{equation}
    \frac{\partial }{\partial {{M}^{2}}}\left[ \int{\frac{{{d}^{3}}p}{{{(2\pi )}^{3}}}{{E}_{p}}} \right]=\int_{-\infty }^{+\infty }{\frac{dx}{2\pi }\int \frac{{{d}^{3}}p}{{{(2\pi )}^{3}}}\frac{1}{{{x}^{2}}+{{E}_{p}}^{2}}}.	
    \end{equation}
For the sake of clarity, we introduce the four-dimensional momentum integration variable $x$. To rigorously separate the vacuum contribution from the integral, we apply the following identity three times consecutively \cite{36, 61}, such that the integrand function can be designed as
	\begin{equation}\frac{1}{{{x}^{2}}+{{E}_{p}}^{2}}=\frac{1}{{{x}^{2}}+{{p}^{2}}+M_{0}^{2}}+\frac{{{p}^{2}}+M_{0}^{2}-{{E}_{p}}^{2}}{({{x}^{2}}+{{p}^{2}}+M_{0}^{2})[{{x}^{2}}+{{E}_{p}}^{2}]},	
    \end{equation}
or
	\begin{equation}
    \begin{aligned}
  & \frac{1}{{{x}^{2}}+{{E}_{p}}^{2}}=\frac{1}{{{x}^{2}}+{{p}^{2}}+M_{0}^{2}}-\frac{{{A}_{s}}(p)}{{{\left( {{x}^{2}}+{{p}^{2}}+M_{0}^{2} \right)}^{2}}} \\
 & +\frac{A_{s}^{2}(p)}{{{\left( {{x}^{2}}+{{p}^{2}}+M_{0}^{2} \right)}^{3}}}-\frac{A_{s}^{3}(p)}{{{\left( {{x}^{2}}+{{p}^{2}}+M_{0}^{2} \right)}^{3}}\left[ {{x}^{2}}+{{E}_{p}}^{2} \right]}
\end{aligned},
\end{equation}
where ${{A}_{s}}(p)=\mu _{5}^{2}+2sp{{\mu }_{5}}+{{M}^{2}}-M_{0}^{2}$, ${{M}_{0}}$ the quark mass in vacuum (i.e., the value calculated at $T=0$ and ${{\mu }_{5}}=0$).  At $T = 0$, the gap equation is given as
	\begin{equation}
    \frac{M-{{m}_{c}}}{2G}-M{{N}_{c}}{{N}_{f}}{{I}_{M}}=0,	
    \end{equation}
where
	\begin{equation}
    {{I}_{M}}=\sum\limits_{s=\pm 1}{\int{\frac{{{d}^{3}}p}{{{(2\pi )}^{3}}}\frac{1}{{{E}_{p}}}}}.	
    \end{equation}

The specific MSS method of ${{I}_{M}}$ can be given as \cite{26, 37, 48}
	\begin{equation}
    {{I}_{M}}=2{{I}_{\text{quad}}}\left( {{M}_{0}} \right)-\left( {{M}^{2}}-M_{0}^{2}-2\mu _{5}^{2} \right){{I}_{\text{log}}}\left( {{M}_{0}} \right)+\left[ \frac{3{{(M_{0}^{2}-{{M}^{2}}-\mu _{5}^{2})}^{2}}}{4} \right]{{I}_{1}}+2{{I}_{2}},	
    \end{equation}
where
	\begin{equation}
    {{I}_{\text{quad}}}({{M}_{0}})=\underset{{}}{\overset{{}}{\mathop{\int_{0}^{\Lambda}{\frac{dp}{2{{\pi }^{2}}}\frac{{{p}^{2}}}{\sqrt{{{p}^{2}}+M_{0}^{2}}}}}}}\,,	
    \end{equation}
	\begin{equation}
    {{I}_{\log }}({{M}_{0}})=\int_{0}^{\Lambda}{\frac{dp}{2{{\pi }^{2}}}\frac{{{p}^{2}}}{{{\left( {{p}^{2}}+M_{0}^{2} \right)}^{3/2}}}},	
    \end{equation}
	\begin{equation}
    {{I}_{1}}=\int_{0}^{\infty }{\frac{\text{d}p}{2{{\pi }^{2}}}\frac{{{p}^{2}}}{{{\left( {{p}^{2}}+M_{0}^{2} \right)}^{5/2}}}},	
    \end{equation}
	\begin{equation}
    \begin{aligned}
   {{I}_{2}}&=\frac{15}{32}\sum\limits_{s=\pm 1}{\int{\frac{{{\text{d}}^{3}}p}{{{(2\pi )}^{3}}}}}\int_{0}^{1}{\text{d}t{{(1-t)}^{2}}} \\
 & \times \frac{{{(M_{0}^{2}-{{M}^{2}}-\mu _{5}^{2}-2s{{\mu }_{5}}\sqrt{{{p}^{2}}+{{M}^{2}}})}^{3}}}{{{[\left( 2s{{\mu }_{5}}\sqrt{{{p}^{2}}+{{M}^{2}}}-M_{0}^{2}+{{M}^{2}}+\mu _{5}^{2} \right)t+{{p}^{2}}+M_{0}^{2}]}^{7/2}}}
\end{aligned},	
\end{equation}
where $\Lambda $ denotes the regularization parameter for divergent integrals, and ${{I}_{\text{quad}}}(\Lambda,{{M}_{0}})$ and ${{I}_{\log }}\left( \Lambda,{{M}_{0}} \right)$ represent the ultraviolet divergent integrals of quadratic divergence and logarithmic divergence, respectively. Notably, the medium contribution has been completely separated from the divergent integrals, leaving only ${{I}_{\text{quad}}}({{M}_{0}})$ and ${{I}_{\log }}({{M}_{0}})$ - the explicit functions of ${{M}_{0}}$ - to require regularization. In contrast, ${{I}_{1}}$ and ${{I}_{2}}$ are both ultraviolet finite and can be directly evaluated by extending the integration region to infinity in the momentum integral. The only difference of the integration region between TRS and MSS lies in the ${{\Omega}_{\text{vac}}}$ term as
 \begin{equation}
  \Omega _V^{{\rm{MSS}}} =  - 2{N_c}{N_f}\left\{ {\frac{{{M^2} - M_0^2}}{2}{I_{{\rm{quad}}}}({M_0})} \right.\left. { + \left[ {\mu _5^2{M^2} + \frac{{{{(M_0^2 - {M^2})}^2}}}{4}} \right]\frac{{{I_{\log }}({M_0})}}{2} + {I_{{\rm{fin}}}}} \right\},
 \end{equation}
   \begin{equation}
{I_{{\rm{fin}}}} = \int\limits_0^\infty  {\frac{{dp\;{p^2}}}{{2{\pi ^2}}}} [\frac{{{{({M^2} - M_0^2)}^2} - 4{M^2}\mu _5^2}}{{8{{({p^2} + {M_0}^2)}^{3/2}}}} - \frac{{{M^2} - M_0^2}}{{2\sqrt {{p^2} + {M_0}^2} }} - \sqrt {{p^2} + {M_0}^2}  + \frac{1}{2}\sum\limits_{s =  \pm 1} {{E_p}} ].
 \end{equation}

An alternative version of the MSS thermodynamic potential may be obtained from the mass gap equation, which after performing all the finite integrations analytically, we obtain
	\begin{equation}
    \begin{aligned}
   \frac{M-m}{4{{N}_{f}}{{N}_{c}}{{G}_{s}}M}&={{I}_{\text{quad}}}(\Lambda,{{M}_{0}})
  +\left( 2\mu _{5}^{2}-{{M}^{2}}+M_{0}^{2} \right){{I}_{\log }}\left( \Lambda,{{M}_{0}} \right)\\
  &-\frac{2\mu _{5}^{2}+{{M}^{2}}-M_{0}^{2}}{8{{\pi }^{2}}}
  +\frac{{{M}^{2}}-2\mu _{5}^{2}}{8{{\pi }^{2}}}\ln \left( \frac{{{M}^{2}}}{M_{0}^{2}} \right)
\end{aligned}.
\end{equation}
Note that the quark mass dependence of both ${{I}_{\text{quad}}}({{M}_{0}})$ and ${{I}_{\log }}({{M}_{0}})$ is only reflected through the vacuum quark mass ${{M}_{0}}$. Integrating Eq. (23) with respect to, we obtain
\begin{equation}
   \Omega _V^{{\rm{MSS}}} =  - 2{N_c}{N_f}\left\{ \begin{array}{l}
\frac{{{M^2}}}{{8{\pi ^2}}}[\frac{{{M^2}}}{4} - \mu _5^2]\ln (\frac{{{M^2}}}{{M_0^2}}) - \frac{{3{M^4}}}{{64{\pi ^2}}} + \frac{{{M^2}M_0^2}}{{16{\pi ^2}}}\\
 + \frac{{{M^2}}}{2}{I_{{\rm{quad}}}}({M_0}) + \left[ {\mu _5^2{M^2} - \frac{{{M^4}}}{4} + \frac{{{M^2}M_0^2}}{2}} \right]\frac{{{I_{\log }}({M_0})}}{2}
\end{array} \right\}.
\end{equation}

\section{NUMERICAL RESULT}
To ensure consistency between the parameters of the NJL model under the rotating framework and experimental observables, we directly use the calibrated results from Ref. \cite{17}: the bare quark mass $m$ = 0.006 GeV, constituent quark mass $M_0$ = 0.302 GeV, cutoff $\Lambda=0.62676\ \text{GeV}$, and the four-fermion coupling strength fixed at ${{G}_{s}}=2.02/{{\Lambda}^{2}}$. For the angular momentum quantum number $n$ about the $z$-axis, although theoretically it should be summed from $n=0,\,\pm 1,\,\pm 2,\,...$ to infinity, numerical results show that the series converges rapidly-keeping $n\in \left[ -5,\,5 \right]$ is enough to ensure a small error \cite{46,62,63}. Additionally, to ensure that the calculations do not exceed the speed of light, the condition $\omega \,r<1$ must always be met throughout the computation.

Figure 1 illustrates how the chiral chemical potential $\mu_5$ and rotation affect the chiral phase transition. In panel (a), it shows that: As $\mu_5$ increases, the quark mass $M$ increases at low temperatures, and the transition to the chiral symmetry restored phase occurs at higher temperatures. This indicates that chiral imbalance enhances chiral symmetry breaking and raises the pseudocritical temperature $T_{\textrm{pc}}$. In panel (b), where rotation is present, the dynamical quark mass decreases more rapidly with temperature compared to panel (a), and the transition occurs at lower temperatures. This shows that rotation weakens chiral symmetry breaking and lowers $T_{\textrm{pc}}$, consistent with centrifugal and spin-orbit coupling effects. It is found that chiral chemical potential strengthens chiral symmetry breaking, while angular momentum weakens it. The suppression of $T_{\textrm{pc}}$ by rotation is less pronounced at higher $\mu_5$, indicating that chiral imbalance can partially counteract the disordering effect of rotation.
\begin{figure}[h]
	\centering
	\includegraphics[width=0.85\textwidth]{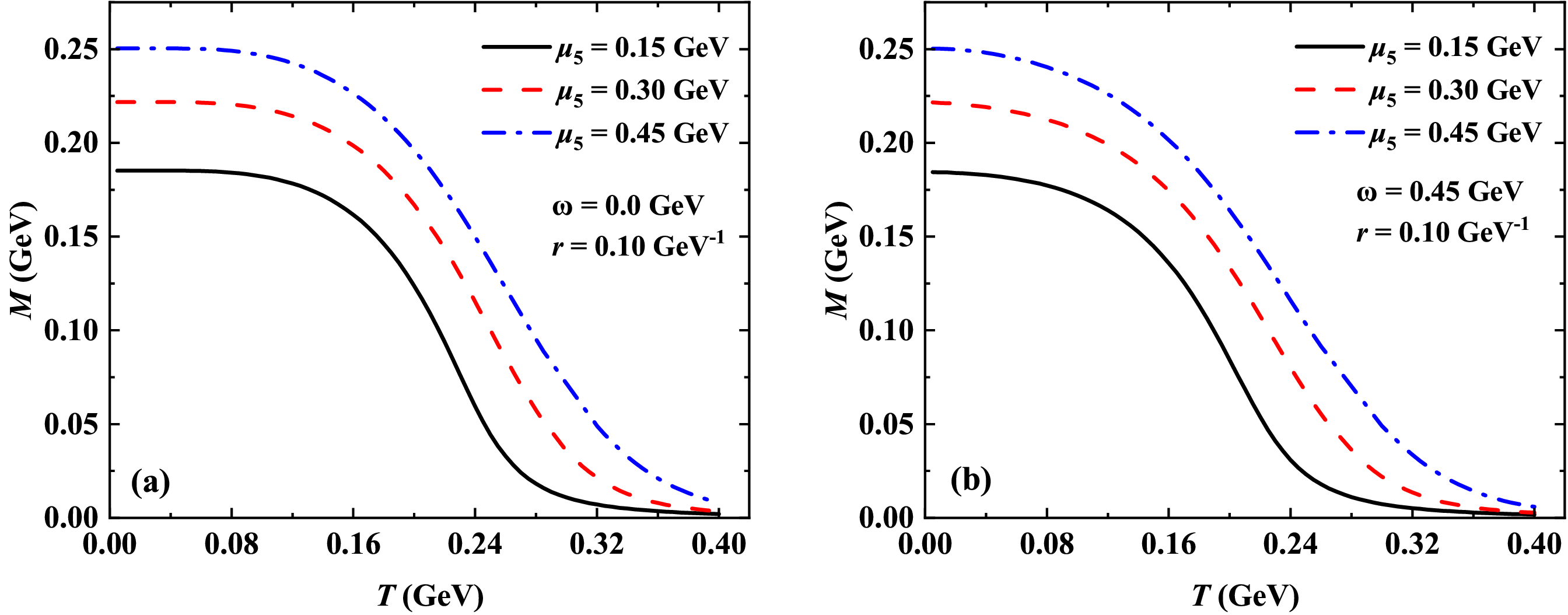}
	\caption{\label{fig1} Temperature dependence of the quark mass $M$ calculated using the MSS, with a fixed rotation radius $r = 0.1\,\text{Ge}{{\text{V}}^{\text{-1}}}$. Panel (a) shows the case of zero angular velocity ($\omega =0\ \text{GeV}$), while panel (b) corresponds to $0.45\ \text{GeV}$. The solid, dashed, and dot-dashed lines represent chiral chemical potentials ${{\mu }_{5}}=0.15\ \text{GeV}$, $0.3\ \text{GeV}$, and $0.45\ \text{GeV}$,  respectively.}
	\label{fig:1}
\end{figure}

Figure 2 demonstrates how rotation and chiral chemical potential jointly influence the chiral symmetry breaking. Comparing panel (a) ($\mu_5$ = 0) and panel (b) ($\mu_5$= 0.15 GeV),  it is evident that the introduction of  $\mu_5$ significantly enhances the quark mass at low temperatures and shifts the transition to higher temperatures. This confirms that chiral imbalance strengthens chiral symmetry breaking and raises $T_{\textrm{pc}}$. In panel (b), although increasing $\omega$ still suppresses $M$ and $T_{\textrm{pc}}$, the overall scale of $M$ remains higher than in panel (a) due to the presence of  $\mu_5$. This demonstrates that chiral imbalance can partially counteract the rotation-induced suppression of chiral symmetry breaking.
\begin{figure}[h]
	\centering
	\includegraphics[width=0.85\textwidth]{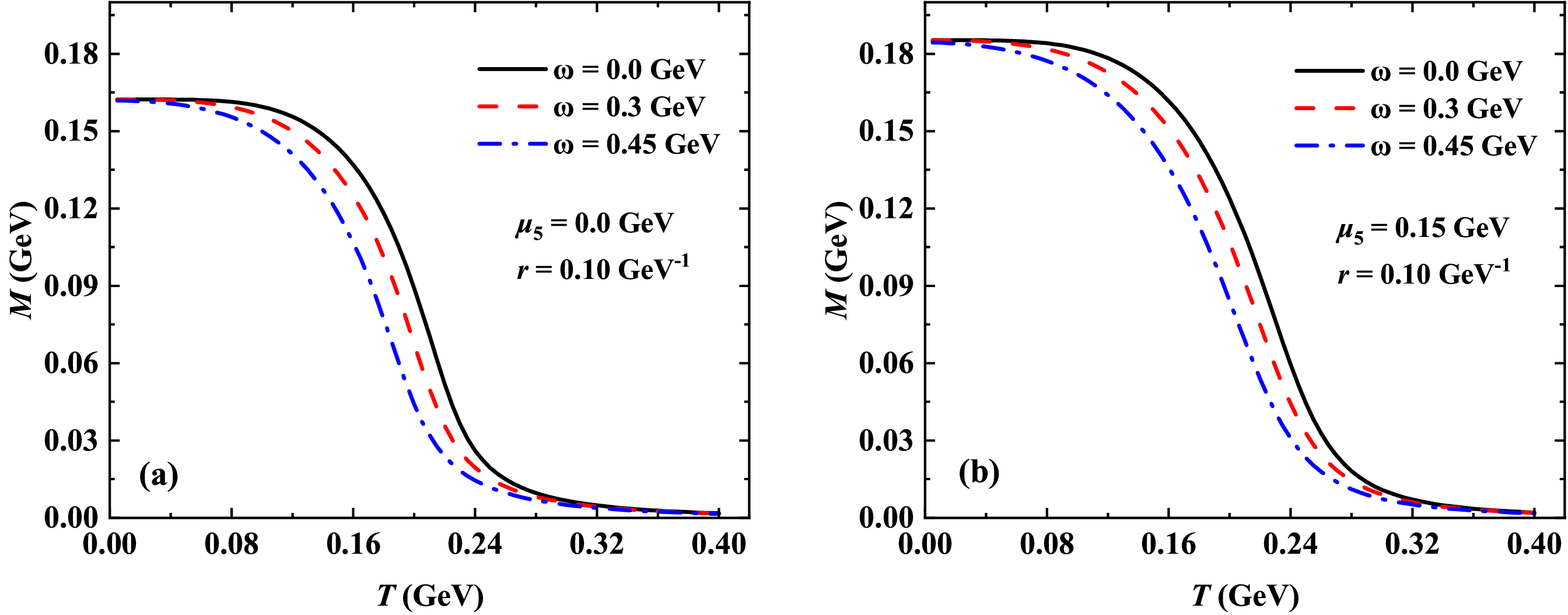}
	\caption{\label{fig2}  Temperature dependence of the quark mass $M$, with a fixed rotation radius $r=0.1\,\text{Ge}{{\text{V}}^{\text{-1}}}$. Panel (a) corresponds to the case of vanishing chiral chemical potential (${{\mu }_{5}}=0\ \text{GeV}$), while panel (b) corresponds to finite chiral chemical potential (${{\mu }_{5}}=0.15\ \text{GeV}$). The solid, dashed, and dot-dashed lines denote angular velocities $\omega =0\ \text{GeV}$, $0.3\ \text{GeV}$, and $0.45\ \text{GeV}$, respectively.}
	\label{fig:2}
\end{figure}

\newpage
Figure 3 illustrates the temperature evolution of the derivative of the quark mass with respect to temperature, d$M$/d$T$, calculated using the Medium Separation Scheme (MSS) within the two-flavor Nambu-Jona-Lasinio (NJL) model. Figure 3 focuses on how rotation and chiral chemical potential affect not only the transition temperature but also the nature (smoothness vs. sharpness) of the chiral phase transition. It reveals that as $\omega$ increases, the peak position shifts systematically toward lower temperatures, confirming that rotation suppresses $T_{\textrm{pc}}$. More importantly, the peak height decreases and the peak width broadens with increasing $\omega$, indicating that rotation softens the phase transition which making it more crossover-like and less abrupt. Comparing panel (a) and panel (b), the introduction of $\mu_5$ shifts the overall peak positions toward higher temperatures, consistent with the enhancement of chiral symmetry breaking by chiral imbalance. The presence of $\mu_5$ compensates for the rotation-induced softening of the phase transition. At finite $\mu_5$, the reduction in peak height and the broadening caused by rotation are less pronounced. This demonstrates that chiral imbalance not only raises $T_{\textrm{pc}}$ but also helps maintain a sharper phase transition even under strong rotation, effectively counteracting the disordering effect of centrifugal forces and spin-orbit coupling.
\begin{figure}[h]
	\centering
	\includegraphics[width=0.85\textwidth]{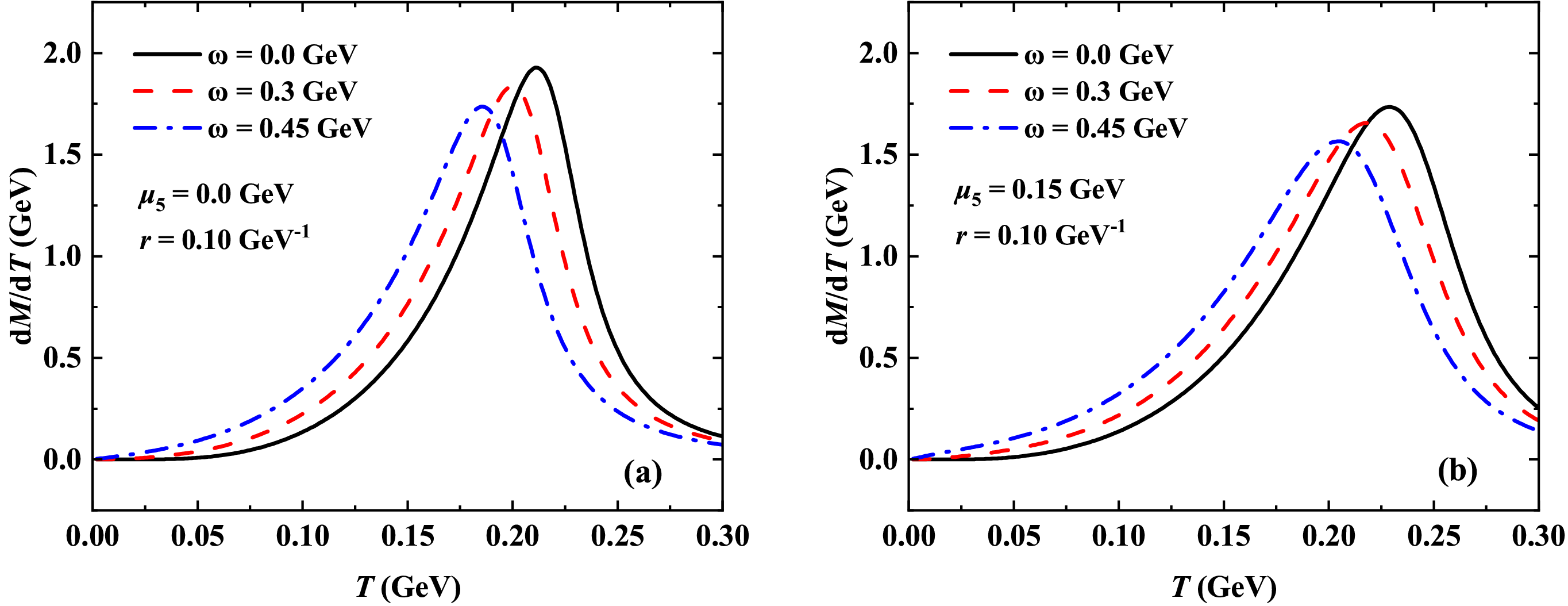}
	\caption{\label{fig3}  Temperature evolution of the derivative of the quark mass with respect to temperature, $dM/dT$. Panel (a) corresponds to vanishing chiral chemical potential (${{\mu }_{5}}=0\ \text{GeV}$), while panel (b) corresponds to ${{\mu }_{5}}=0.15\ \text{GeV}$. The solid, dashed, and dot-dashed lines represent angular velocities $\omega =0\ \text{GeV}$, 0.3 GeV, and 0.45 GeV, respectively.}
	\label{fig:3}
\end{figure}

\newpage
Figure 4 presents the evolution of the pseudocritical temperature $T_{pc}$ of the chiral phase transition as a function of the chiral chemical potential $\mu_5$. Figure 4 focuses on the combined effects of chiral chemical potential and rotation on the chiral phase transition temperature. It reveals that $T_{\textrm{pc}}$ exhibits a monotonically increasing behavior with increasing $\mu_5$ for all angular velocities considered. This indicates that chiral imbalance enhances chiral symmetry breaking and raises the critical temperature of the phase transition. At any fixed $\mu_5$, increasing $\omega$ systematically
suppresses $T_{\textrm{pc}}$, confirming that rotation weakens chiral symmetry breaking. The suppression is evident from the downward shift of the curves as $\omega$ increases. Importantly, the suppression of  $T_{\textrm{pc}}$  by rotation becomes gradually weaker as $\mu_5$ increases. At small $\mu_5$, the three curves are well separated, indicating strong rotational suppression. At large $\mu_5$, the curves converge, showing that chiral imbalance counteracts the destructive effect of rotation. This demonstrates that chiral imbalance helps the system maintain a relatively high phase transition temperature even under strong rotation.

Previous studies of NJL-type models using TRS predicted that $T_{\textrm{pc}}$ decreases with increasing $\mu_5$, which qualitatively contradicts lattice QCD (LQCD) results \cite{34, 35}. In contrast, MSS predicts a monotonic increase of $T_{\textrm{pc}}$ with $\mu_5$, in qualitative agreement with LQCD.
\begin{figure}[h]
	\centering
	\includegraphics[width=0.45\textwidth]{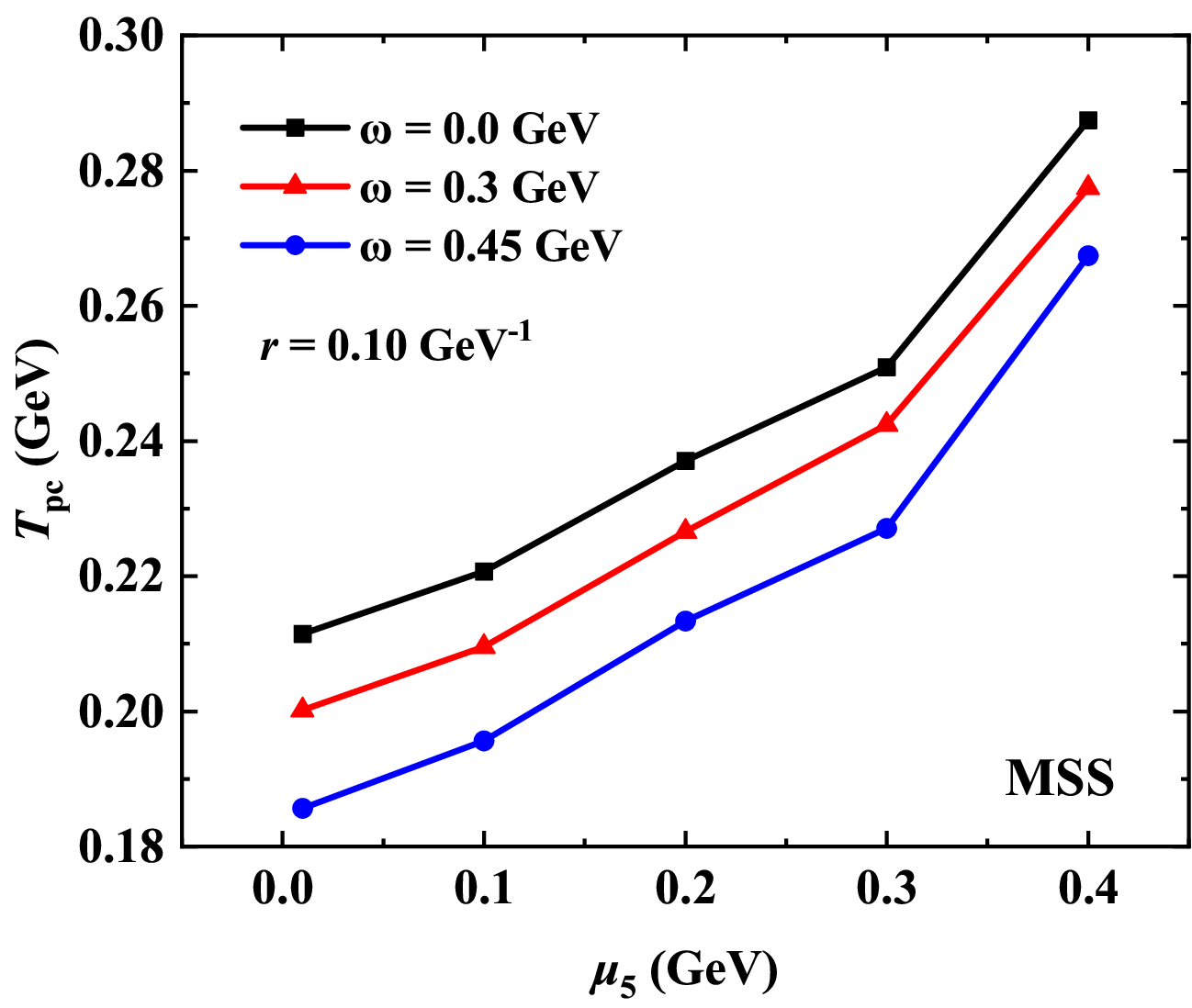}
	\caption{\label{fig4}  Pseudocritical temperature ${{T}_{\textrm{pc}}}$ of the chiral phase transition as a function of the chiral chemical potential ${{\mu }_{5}}$, calculated within the MSS at a fixed rotation radius. Three angular velocities are employed: $\omega =0\ \text{GeV}$ (black squares), $\omega =0.3\ \text{GeV}$ (red triangles), and $\omega =0.45\ \text{GeV}$ (blue circles).}
	\label{fig:4}
\end{figure}

\newpage
Figure 5 illustrates the evolution of the pseudocritical temperature $T_{\textrm{pc}}$ of the chiral phase transition as a function of the angular velocity $\omega$. Figure 5 focuses on how rotation affects the chiral phase transition temperature and how this effect is modulated by chiral imbalance. It reveals that $T_{\textrm{pc}}$ decreases monotonically with increasing $\omega$ for all values of $\mu_5$. This confirms that rotation suppresses chiral symmetry breaking and lowers the critical temperature, consistent with centrifugal effects and spin-orbit coupling. The decreasing trend exhibits a distinct nonlinear behavior: In the low-rotation region ($\omega <$ 0.1 GeV), the decrease of $T_{pc}$ is relatively gentle. In the high-rotation region ($\omega >$ 0.1 GeV), the slope becomes significantly steeper, indicating that the binding energy of quark-antiquark pairs is more drastically weakened under strong rotation. At any fixed $\omega$, $T_{\textrm{pc}}$ is higher for larger $\mu_5$, confirming that chiral imbalance enhances chiral symmetry breaking. More importantly, the suppression of $T_{\textrm{pc}}$ by rotation becomes gradually weaker as $\mu_5$ increases. The curves for different $\mu_5$ are well separated at low $\omega$, but the relative reduction caused by rotation is smaller at higher $\mu_5$. This demonstrates that chiral imbalance buffers the system against rotation-induced suppression of the phase transition temperature.
\begin{figure}[h]
	\centering
	\includegraphics[width=0.45\textwidth]{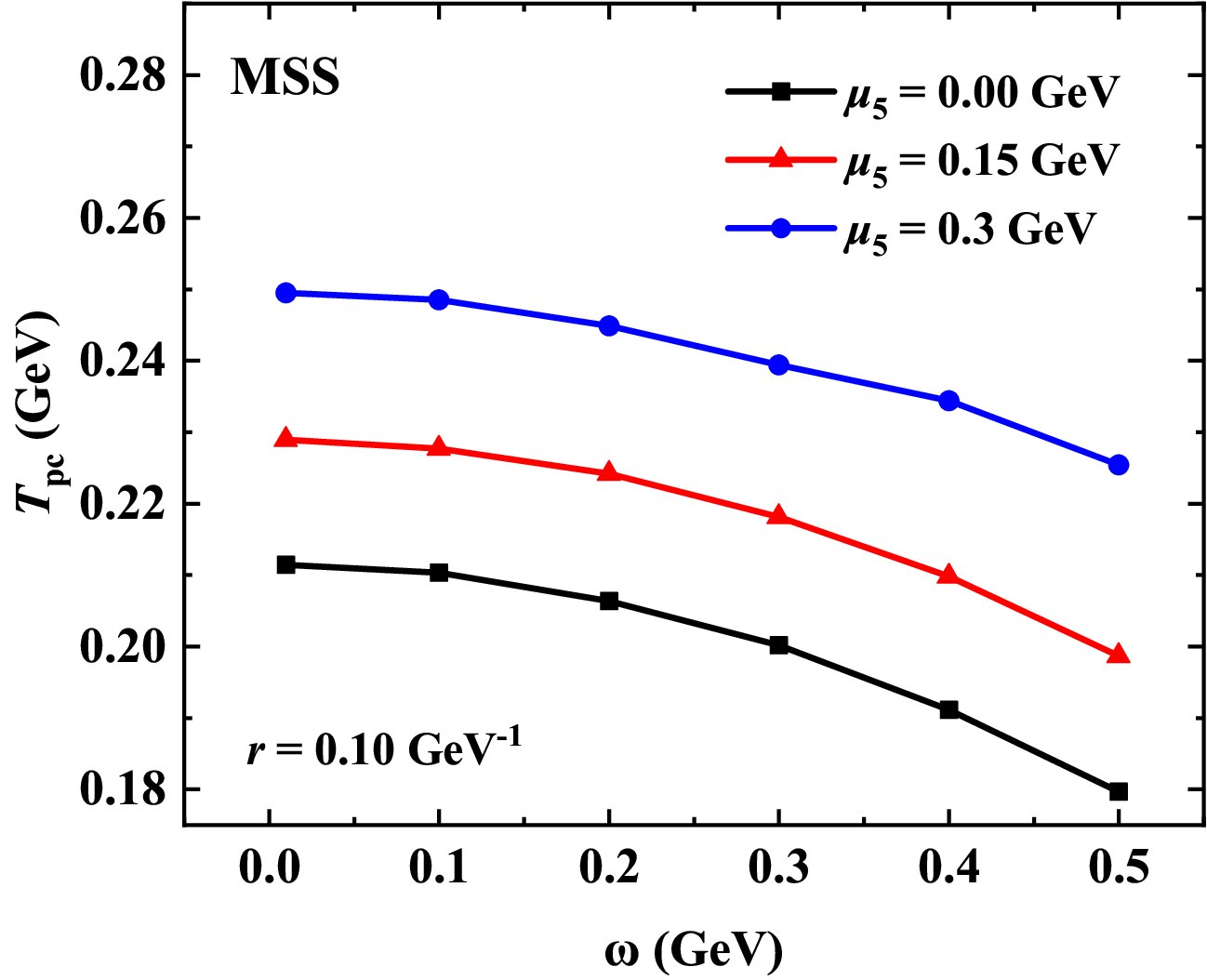}
	\caption{\label{fig5}  Pseudocritical temperature ${{T}_{\textrm{pc}}}$ of the chiral phase transition as a function of angular velocity $\omega $. The black squares, red triangles, and blue circles represent chiral chemical potentials ${{\mu }_{5}}=$ 0.0 GeV, 0.15 GeV, and 0.3 GeV, respectively. }
	\label{fig:5}
\end{figure}

\newpage
Figure 6 focuses on how the rotation radius influences the suppression of the chiral phase transition temperature by rotation. It reveals that for all radii,  $T_{\textrm{pc}}$ decreases monotonically with increasing $\omega$, confirming that rotation universally suppresses chiral symmetry breaking. The suppression effect is highly sensitive to the rotation radius: For a small radius ($r$ = 0.1 GeV$^{-1}$), the decrease is gentle, and  $T_{\textrm{pc}}$ remains at a relatively high level across the entire $\omega$ range. As the radius increases to $r$ = 0.5 GeV$^{-1}$, the slope becomes steeper, indicating a stronger suppression. For the largest radius ($r$ = 1.0 GeV$^{-1}$), the suppression is most dramatic: $T_{\textrm{pc}}$ drops sharply and even exhibits an abrupt drop in the high-rotation region ($\omega >$ 0.15 GeV). A larger rotation radius amplifies the effects of spacetime curvature and centrifugal stretching induced by rotation.

The influence of rotation on the chiral phase transition is not uniform, which means it depends crucially on the distance from the rotation axis. This highlights the importance of considering the spatial structure of rotating systems. Larger radii experience stronger centrifugal forces and greater spacetime curvature, which more effectively tear apart quark-antiquark bound states. The sudden decrease in  $T_{\textrm{pc}}$  for large $r$ and high $\omega$ may indicate a phase transition weakening or even a potential critical behavior in the rotational response of the QCD vacuum. In non-central collisions, the produced QGP has a spatially varying angular velocity profile. Figure 6 suggests that regions farther from the center experience a stronger suppression of chiral symmetry breaking, leading to spatial inhomogeneity in the phase transition temperature.
\begin{figure}[h]
	\centering
	\includegraphics[width=0.45\textwidth]{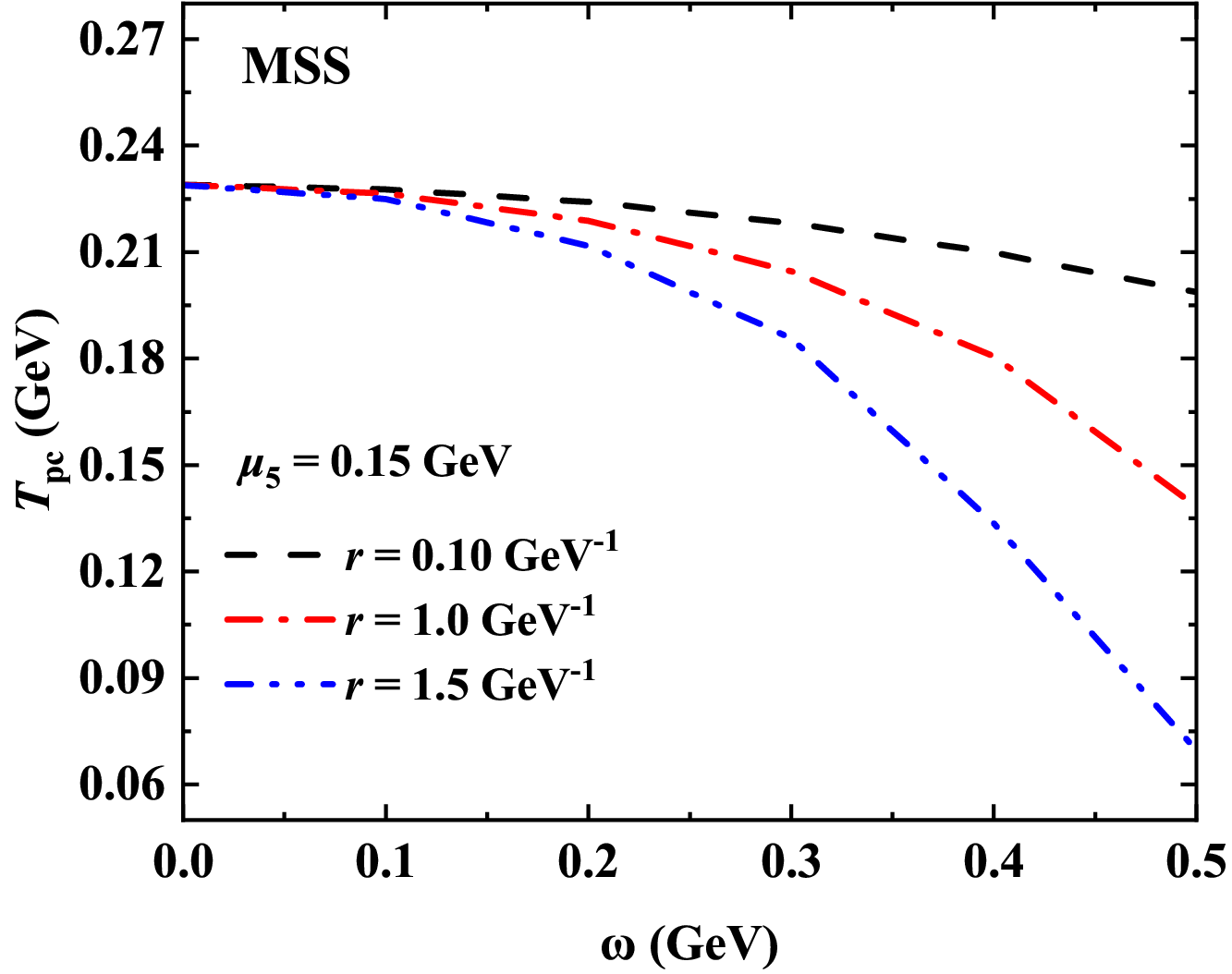}
	\caption{\label{fig6}  Pseudocritical temperature ${{T}_{\text{pc}}}$ of the chiral phase transition as a function of angular velocity $\omega $. The black dashed, red dash-dotted, and blue dotted lines represent rotation radii $r = 0.1\,\text{Ge}{{\text{V}}^{-1}}$, $r=1\ \text{Ge}{{\text{V}}^{-1}}$, and $r=1.5\,\text{Ge}{{\text{V}}^{-1}}$, respectively.}
	\label{fig:6}
\end{figure}

\section{SUMMARY AND CONCLUSIONS}
In this work, we adopt the Medium Separation Scheme (MSS) to regularize the two-flavor Nambu-Jona-Lasinio (NJL) model and systematically investigate the chiral phase transition in rotating quark matter with chiral imbalance. The competing and cooperative interplay between rotation and the chiral chemical potential $\mu_5$ is elucidated.

Our numerical results reveal that $\mu_5$ and rotation exert opposing influences on chiral symmetry breaking: increasing $\mu_5$ enhances chiral symmetry breaking, raises the pseudocritical temperature $T_{\textrm{pc}}$, and sharpens the phase transition; whereas increasing the angular velocity $\omega$ suppresses chiral symmetry breaking through centrifugal and spacetime curvature effects, leading to a monotonic decrease in $T_{\textrm{pc}}$ and a smearing of the transition. A key finding is that chiral imbalance provides a significant buffering effect against rotation-induced suppression of the phase transition: the reduction of $T_{\textrm{pc}}$ with increasing $\omega$ becomes progressively weaker as $\mu_5$ grows, indicating that enhanced quark spin polarization and chiral pairing help stabilize the chiral condensate against the disordering effects of rotation.

Notably, within the MSS framework, $T_{\textrm{pc}}$ exhibits a monotonic increase with $\mu_5$, in qualitative agreement with lattice QCD results, thereby resolving the long-standing discrepancy found in traditional regularization schemes. Furthermore, the suppression of $T_{\textrm{pc}}$ by rotation is found to be strongly radius-dependent: larger rotation radii amplify the suppression due to enhanced spacetime curvature and centrifugal stretching, and can even induce an abrupt drop in $T_{\textrm{pc}}$ in the high-rotation region.

This work establishes the applicability of the MSS as a reliable regularization approach for describing rotating chiral-imbalanced quark matter and uncovers the competitive yet cooperative roles of rotation and chiral imbalance in modulating the QCD chiral phase transition. These findings provide valuable theoretical insights into the evolution of strongly interacting matter under extreme rotational conditions, such as those encountered in non-central heavy-ion collisions. Future work will extend this analysis to the three-flavor NJL model and incorporate isospin asymmetry and external magnetic fields to explore multi-field modulation of the chiral phase transition.

\begin{acknowledgments}
	This work was supported by the National Natural Science Foundation of China (Grants No. 12575144, and No. 11875178).
\end{acknowledgments}

\bibliography{Feng}

\end{document}